\documentclass[conference]{IEEEtran}
\IEEEoverridecommandlockouts
\usepackage{cite}
\usepackage{amsmath,amssymb,amsfonts}
\usepackage{algorithmic}
\usepackage{graphicx}
\usepackage{textcomp}
\usepackage{xcolor}
\usepackage{booktabs}
\usepackage{hyperref}
\usepackage[utf8]{inputenc}
\usepackage{subcaption}
\def\BibTeX{{\rm B\kern-.05em{\sc i\kern-.025em b}\kern-.08em
    T\kern-.1667em\lower.7ex\hbox{E}\kern-.125emX}}
\begin{document}

\title{Diffuse or Confuse: A Diffusion Deepfake Speech Dataset}

\author{\IEEEauthorblockN{1\textsuperscript{st} Anton Firc}
\IEEEauthorblockA{\textit{Faculty of Information Technology} \\
\textit{Brno University of Technology}\\
Brno, Czech Republic \\
ifirc@fit.vut.cz}
\and
\IEEEauthorblockN{2\textsuperscript{nd} Kamil Malinka}
\IEEEauthorblockA{\textit{Faculty of Information Technology} \\
\textit{Brno University of Technology}\\
Brno, Czech Republic \\
malinka@fit.vut.cz}
\and
\IEEEauthorblockN{3\textsuperscript{rd} Petr Hanáček}
\IEEEauthorblockA{\textit{Faculty of Information Technology} \\
\textit{Brno University of Technology}\\
Brno, Czech Republic \\
hanacek@fit.vut.cz}
}

\maketitle

\begin{abstract}
Advancements in artificial intelligence and machine learning have significantly improved synthetic speech generation. This paper explores diffusion models, a novel method for creating realistic synthetic speech. We create a diffusion dataset using available tools and pretrained models. Additionally, this study assesses the quality of diffusion-generated deepfakes versus non-diffusion ones and their potential threat to current deepfake detection systems. Findings indicate that the detection of diffusion-based deepfakes is generally comparable to non-diffusion deepfakes, with some variability based on detector architecture. Re-vocoding with diffusion vocoders shows minimal impact, and the overall speech quality is comparable to non-diffusion methods.
\end{abstract}


\section{Introduction}

Advancements in artificial intelligence and machine learning have significantly improved synthetic speech generation~\cite{FIRC2023e15090}. The quality of deepfake speech is now advanced enough to deceive both systems~\cite{10.1145/3477314.3507013} and humans~\cite{10346006}. Recently, diffusion models have emerged as a new technique for producing highly realistic synthetic speech~\cite{zhang2023survey}. This paper examines the creation of a diffusion dataset using available tools and pretrained models and provides a detailed comparison of these models' speech properties and their detectability by deepfake speech detection methods, in contrast to traditional methods.

Unlike traditional GANs, diffusion models iteratively refine data through a reverse diffusion process, promising more natural and convincing speech. This could pose new challenges for deepfake detection systems~\cite{diffusion_models}.

Our study investigates critical aspects of diffusion-based synthetic speech. We compare diffusion-generated deepfakes to non-diffusion-generated ones, evaluating the quality and characteristics of the synthetic speech to determine if they present a greater threat to current detection algorithms. Additionally, we assess whether training on a single non-diffusion dataset (ASVspoof2019 LA~\cite{WANG2020101114}) is sufficient for effectively detecting dif\-fu\-sion-based deepfakes.

The primary contributions of this paper are as follows:
\begin{itemize}
    \item We release a diffusion-generated deepfake speech dataset to facilitate further research and development in synthetic speech and deepfake detection.
    \item We conduct a preliminary assessment of the impact of diffusion-generated speech on existing deepfake detection systems. 
    \item We compare the speech quality produced by diffusion models to \textit{traditional}, non-diffusion methods.
\end{itemize}

The dataset and supplementary material are available at \texttt{\url{https://github.com/AntonFirc/diffusion-deepfake-speech-dataset/}}.

\section{Related Work}

\begin{figure}[ht]
    \centering
    \includegraphics[width=\linewidth]{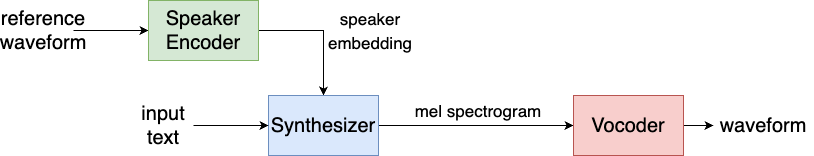}
    \caption{General zero-shot speech synthesis pipeline~\cite{sv2tts}.}
    \label{fig:speech-synth-pipeline}
\end{figure}

Text-to-speech synthesis (TTS) converts text into human-like speech. Modern systems generate speech in various voices, including unseen speakers~\cite{sv2tts}. The ability to generate speech for previously unseen speakers is known as a \textit{zero-shot} setting. In this scenario, the target speech is adapted using only a short embedding utterance. The general zero-shot TTS pipeline involves a speaker encoder, synthesizer, and vocoder, as illustrated in Fig.~\ref{fig:speech-synth-pipeline}. \textit{Speaker encoder} extracts speaker-specific embeddings from the reference waveform. \textit{Synthesizer} converts the input text into a mel spectrogram while being conditioned by the speaker embedding to represent the speaker from the reference waveform. Finally, \textit{vocoder} converts the mel spectrogram into a raw waveform.

Traditional non-diffusion models use generative architectures such as Tacotron2~\cite{tacotron2}, which converts text to mel spectrograms using a sequence-to-sequence architecture. Improvements like location-relative attention enhance alignment and naturalness of Tactotron2 synthesized speech~\cite{tacotron2DCA}. GlowTTS~\cite{glowTTS} autonomously aligns text and speech, speeding up synthesis while supporting multi-speaker scenarios. FastPitch~\cite{fastpitch} supports parallel mel-spectrogram generation, which is achieved by eliminating the auto-regressive loop during inference. Additionally, it enhances prosody by adding pitch prediction. VITS \cite{pmlr-v139-kim21f} combines VAEs with normalizing flows for high-quality end-to-end synthesis and fine-grained control.

Diffusion models, a new generative approach, gradually add and remove noise to generate data~\cite{diffusion_models}. Traditional generative models often rely on direct mapping from a noise distribution to the target data in one step, such as GANs or VAEs. In contrast, diffusion models involve a multi-step process. During the forward process, noise is added incrementally to the data until it's completely corrupted. The reverse process then denoises the data step by step, reconstructing the original data. This iterative noise-adding and denoising mechanism allows diffusion models to generate high-quality and more diverse data than traditional models, making them particularly useful in high-resolution image generation and speech synthesis~\cite{zhang2023survey}.

Diffusion synthesizers use diffusion models to predict the Mel-spectrogram. The spectrogram then may be vocoded using GAN or diffusion vocoder. DiffGAN-TTS~\cite{liu2022diffgantts} combines denoising diffusion probabilistic models (DDPMs) with generative adversarial networks (GANs) to create an efficient text-to-speech system. In this model, the DDPM serves as the generator, refining speech samples through denoising steps, while the GAN framework enhances denoising accuracy through adversarial training. This integration enables rapid, high-quality speech synthesis with fewer denoising steps, making the DDPM suitable for real-time use. DiffSpeech~\cite{liu2021diffsinger} is a neural text-to-speech model utilizing diffusion probabilistic models for high-quality speech generation. Unlike autoregressive models, DiffSpeech uses a non-autoregressive architecture, allowing parallel generation of speech frames and faster synthesis. ProDiff~\cite{huang2022prodiff} is a progressive fast diffusion model designed for high-quality text-to-speech synthesis. Unlike previous models that estimate the gradient for data density and require hundreds or thousands of iterations, ProDiff parameterizes the denoising model by directly predicting clean data, avoiding quality degradation during accelerated sampling. Grad-TTS~\cite{pmlr-v139-popov21a} gradually transforms noise predicted by the encoder and aligns it with text input using a Monotonic Alignment Search. Grad-TTS extends conventional diffusion probabilistic models by employing stochastic differential equations to reconstruct data from noise with varying parameters.

Diffusion vocoders convert Mel-spectrograms to speech using diffusion. WaveGrad~\cite{chen2020wavegrad} is a conditional waveform generation model that estimates data density gradients. Building on score matching and diffusion probabilistic models, WaveGrad starts from a Gaussian white noise signal and iteratively refines it using a gradient-based sampler conditioned on the mel-spectrogram. WaveGrad2~\cite{chen21p_interspeech} estimates the gradient of the log conditional density of the waveform given a phoneme sequence, enabling direct audio generation from phonemes. The model consists of an encoder that extracts abstract hidden representations from the phoneme input, a resampling layer that adjusts the resolution of these representations to match the output waveform's time scale, and a WaveGrad decoder that iteratively refines the noisy waveform to generate the final audio. DiffWave~\cite{kong2021diffwave} is designed for both conditional and unconditional waveform generation. This non-autoregressive model transforms white noise into a structured waveform through a Markov chain with many synthesis steps. DiffWave is efficiently trained by optimizing a variant of the variational bound on data likelihood. BDDM~\cite{lam2022bddm}, or Bilateral Denoising Diffusion Model, parameterizes both the forward and reverse processes with a schedule network and a score network, trained using a novel bilateral modelling objective. This objective achieves a lower bound of the log marginal likelihood that is tighter than conventional surrogates.

Finally, no currently available deepfake speech datasets contain diffusion-synthesized speech. One of the most relevant deepfake speech datasets is the ASVSpoof 2019 database~\cite{WANG2020101114} that contains 12,482 bonafide and 109,978 deepfake samples synthesized using 14 different tools. 

In this paper, we work with the LJSpeech dataset~\cite{ljspeech17}. This dataset comprises 13,100 bonafide short audio clips of a single speaker reading passages from 7 non-fiction books. The total length is almost 24 hours.

\section{Experiment Design}

The primary objective of the experimental part is to evaluate the quality and detectability of speech synthesized using diffusion-based models compared to traditional synthesizers. This assessment will determine if diffusion-based synthesizers produce higher-quality samples and present greater challenges to current deepfake speech detection methods.

We first generate a novel speech database using diffusion-based synthesizers and non-diffusion synthesizers. The non-diffusion deepfakes serve as a baseline for experiments. Both sets of synthesized data are derived from the LJSpeech dataset, which contains 13,100 samples from a single female speaker in English~\cite{ljspeech17}. We use these synthesised datasets to evaluate three state-of-the-art (SOTA) deepfake speech detectors trained on the ASVSpoof2019 database. We compare the Equal Error Rates (EER) to assess if diffusion-generated speech is more challenging for current detection methods.

Most diffusion-based synthesizers use diffusion for Mel-spectrogram generation and GAN-based vocoders for speech conversion. The key difference thus lies in the generated spectrogram. To further investigate, we re-vocode Tacotron2-DCA samples using diffusion-based vocoders and evaluate these with the SOTA detectors to observe any changes in detection performance.

Additionally, we assess the quality of the synthesized speech using metrics such as Word Error Rate, Perceptual Evaluation of Speech Quality, speaker similarity, and Signal-to-Noise Ratio to explore potential improvements of diffusion-based models over traditional ones.

\subsection{Used synthesizers}

We collected the most recent diffusion-based synthesizers with published code and pretrained models. We divide the used synthesizers into four groups: diffusion synthesizers with non-diffusion vocoders, diffusion-only synthesizers, diffusion-based vocoders and non-diffusion synthesizers. Table~\ref{tab:synthesizerList} shows the overview of used synthesisers. All synthesizers are provided with pretrained models for the LJspeech dataset. 

\begin{table}
    \centering
    \begin{tabular}{@{}llll@{}}
    \toprule
        \textbf{Model} & \textbf{Synthesizer} & \textbf{Vocoder} & \textbf{Year} \\ \midrule 
        DiffGAN-TTS~\cite{liu2022diffgantts} & Diffusion & HiFi-GAN & 2022  \\ 
        DiffSpeech~\cite{liu2021diffsinger} & Diffusion & HiFi-GAN & 2021 \\
        ProDiff~\cite{huang2022prodiff} & Diffusion & HiFi-GAN & 2022 \\
        Grad-TTS~\cite{pmlr-v139-popov21a} & Diffusion & HiFi-GAN & 2021  \\ \midrule 
        WaveGrad2~\cite{chen21p_interspeech} & Diffusion & Diffusion & 2021  \\ \midrule
        WaveGrad~\cite{chen2020wavegrad} & N/A & Diffusion & 2020  \\
        BDDM~\cite{lam2022bddm} & N/A & Diffusion & 2022  \\
        DiffWave~\cite{kong2021diffwave} & N/A & Diffusion & 2021  \\ \midrule
        Tacotron2-DCA~\cite{tacotron2DCA} & RNN & MelGAN & 2020  \\
        GlowTTS~\cite{glowTTS} & Encoder-Decoder & WaveGlow & 2020  \\ 
        FastPitch~\cite{fastpitch} & Transformer & WaveGlow & 2021  \\
        VITS~\cite{pmlr-v139-kim21f} & Variational Autoencoder & HiFi-GAN & 2021 \\ \bottomrule
    \end{tabular}
    \caption{Used synthesizers. N/A in \textbf{Synthesizer} column denotes vocoder-only models.}
    \label{tab:synthesizerList}
\end{table}

\subsection{Quality assessment}

To assess the quality of synthesized speech, we use four metrics. Word Error Rate (WER) measures the accuracy of the transcribed speech by comparing it to the reference text, with lower WER indicating higher accuracy. In ideal conditions, the WER is zero, as all the words from the input text are correctly synthesized and recognised. We used \textit{JiWER}\footnote{\url{https://pypi.org/project/jiwer/}} for WER calculation. The  Speaker similarity evaluates how similar the speakers in two recordings are, using the SpeechBrain speaker recognition module with the \textit{ecapa-voxceleb} model. Perceptual Evaluation of Speech Quality (PESQ) assesses the perceived quality of speech by comparing it to a reference signal, providing a score that reflects the clarity and naturalness of the audio. Higher PESQ values indicate better quality. \textit{Pypesq}\footnote{\url{https://pypi.org/project/pypesq/}} library was used. A corresponding bonafide recording from the LJSpeech dataset was used as a reference to calculate the PESQ score, ensuring the content was identical for an accurate comparison. Finally, Signal-to-Noise Ratio (SNR) measures the desired signal level relative to the background noise, with a higher SNR indicating cleaner audio. The SNR calculation is based on WADA-SNR~\cite{kim08e_interspeech}.

In addition to quality metrics, we measure the speed of the synthesis process using the Real Time Factor (RTF). RTF measures the processing speed as \textit{processing-time} / \textit{length-of-audio}.

\subsection{Used Deepfake Speech Detectors}

We evaluate the detectability of diffusion-synthesized speech using three distinct deepfake speech detection methods. These methods span a broad range of current detection strategies:

LFCC-LCNN~\cite{wang21fa_interspeech}: This traditional approach uses a Light Convolutional Neural Network (LCNN) with Linear Frequency Cepstral Coefficients (LFCC) as input features. It serves as a benchmark, widely recognized from the ASVspoof 2019 challenge, by extracting frequency characteristics to distinguish between genuine and fake speech.

Wav2vec + GAT~\cite{tak2022}: This state-of-the-art method integrates Wav2vec, a self-supervised learning model, with Graph Attention Networks (GAT). Wav2vec extracts robust audio features, while GAT enhances feature representation and classification, showcasing the potential of advanced feature extraction combined with powerful classification techniques.

IDSD~\cite{firc24_spectrogram}: This method transforms audio signals into Short-Time Fourier Transform (STFT) spectrograms, processed by a Temporal Convolutional Network (TCN). The TCN detects anomalies and temporal dependencies in these image-like spectrograms.

\section{Results}

The dataset ultimately consists of 14 sets of synthesized speech. The set refers to a synthetic copy of the LJSpeech dataset. These include variations to the DiffGAN-TTS model, namely \textit{aux}, \textit{shallow} and \textit{naive}. The dataset consists of 183,400 deepfake recordings from one English female speaker. 131,000 recordings are synthesized by diffusion-based tools. The total length of the dataset is approximately 336 hours. Metadata includes model setup, quality assessment, detection results and transcriptions.

\subsection{Detection performance}

\begin{figure*}[ht]
    \centering
    \includegraphics[width=0.8\linewidth]{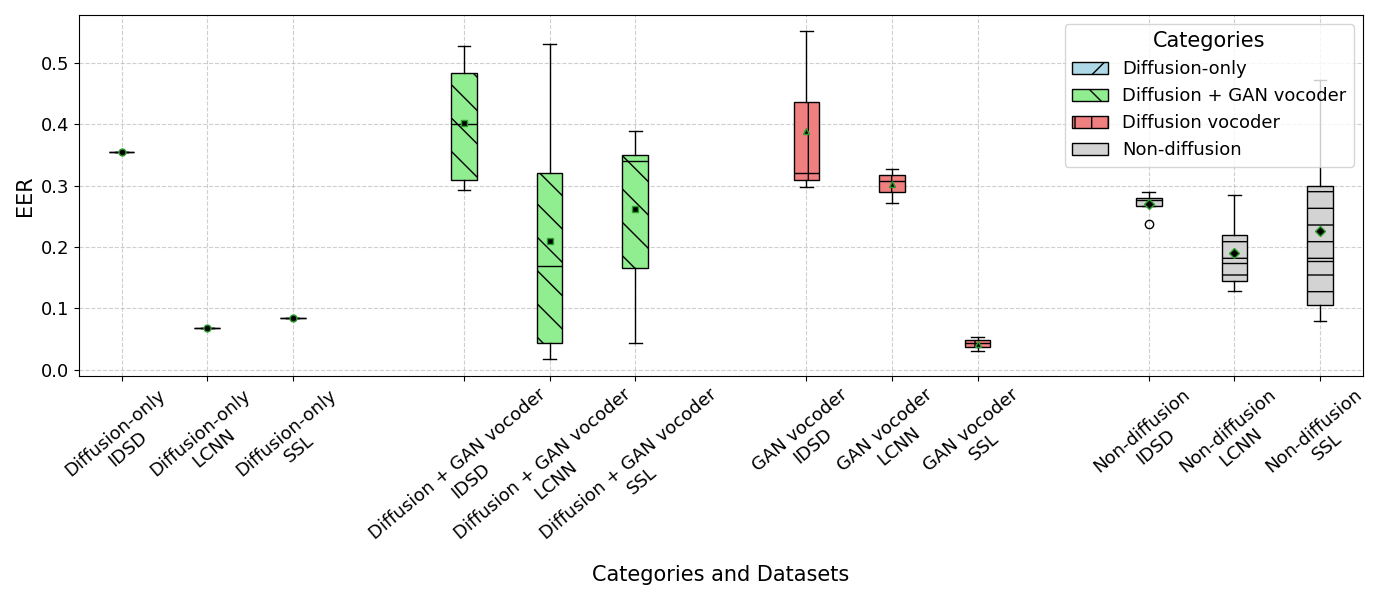}
    \caption{Boxplot visualization of observed EER rates from detectors' validation by category of synthesized speech.}
    \label{fig:eer-boxplot}
\end{figure*}

Each synthesized set was combined with the original LJSpeech data and used as an evaluation set for the employed deepfake speech detectors. The detectors were trained using the ASVSpoof2019 LA training set, which contains only non-diffusion samples. Each evaluation thus consisted of 26,200 samples, split equally between \textit{bonafide} and \textit{deepfake} samples. As Fig.~\ref{fig:eer-boxplot} shows, the resulting EER values are well-distributed in the lower 50\%.

\begin{figure*}[ht]
    \centering
    \includegraphics[width=0.8\linewidth]{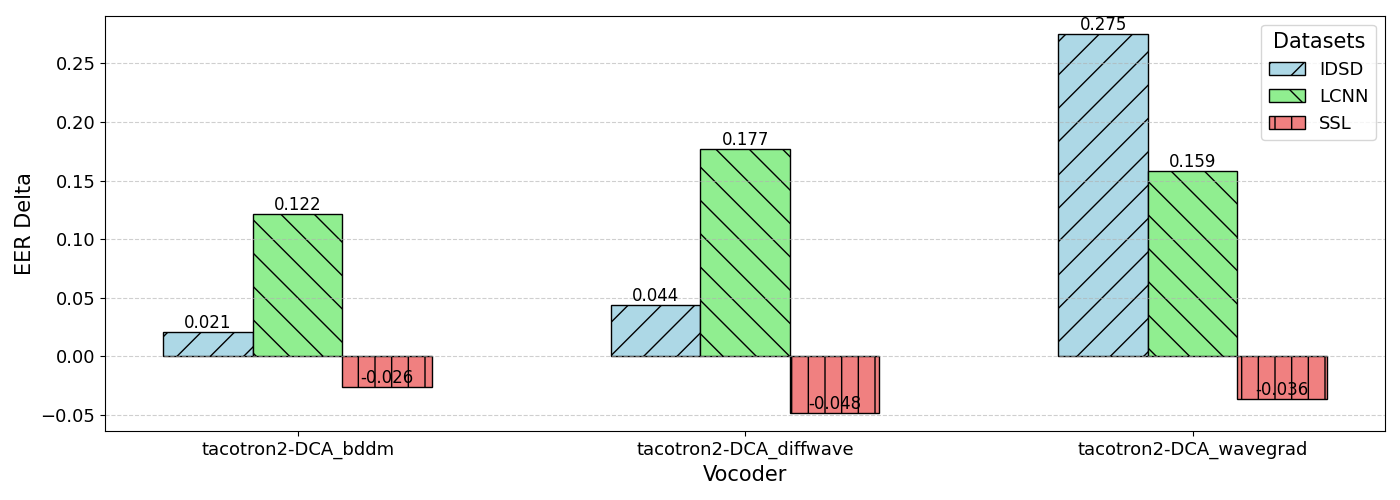}
    \caption{Delta of Equal Error Rates (EER) for detecting non-diffusion Tacotron2-DCA samples compared to re-vocoded Tacotron2-DCA samples using diffusion vocoders.}
    \label{fig:tacotronEER}
\end{figure*}

For the image-based IDSD method, we observe a steady increase in EER when presented with any diffusion-based deepfakes compared to non-diffusion deepfakes. This may be due to the method's direct operation on the spectrogram, where both diffusion synthesizers and vocoders introduce specific artefacts distinct from those produced by non-diffusion algorithms.

The LCNN-based detector generally performs better on diffusion-only synthesized data. This is noteworthy as other technologies introduce more confusion. DiffGAN-TTS synthesized samples show the lowest EER values, indicating that the LCNN-based detector performs well with these samples. This could be due to adversarial training and shallow diffusion steps in DiffGAN-TTS, which result in artefacts that the LCNN network can easily detect. In contrast, the network struggles to detect DiffSpeech, ProDiff, and GradTTS data. These models use more robust and effective denoising processes and optimized diffusion steps, which may produce fewer artefacts, making the synthesized speech more challenging to distinguish from bonafide.

Finally, the SSL-based model shows the most variability in performance over non-diffusion data. The VITS model, employed with the HiFi-GAN vocoder, is particularly problematic, with an observed EER of more than 47\%. Unlike the LCNN network, the SSL-based detector struggles with DiffGAN-TTS-generated samples, likely due to the more varied high-fidelity artefacts in these samples. 

\subsection{Re-vocoding non-diffusion deepfakes}

To assess the impact of re-vocoding non-diffusion generated deepfakes with diffusion vocoders, we took the spectrogram output of the tacotron2-DCA synthesizer. We then used the diffusion-based vocoders to convert the spectrograms to speech. This way, it is possible to assess if re-vocoding non-diffusion deepfake samples makes them more challenging to detect. As shown in Fig.~\ref{fig:tacotronEER} the LCNN network experiences the most noticeable performance deterioration with re-vocoding the Tacotron2 samples. IDSD method struggles, but the difference between Tacotron2 and BDDM and DiffWave vocoded data is not that significant. In contrast, the SSL-based detector performs better on the diffusion-vocoded data. This contrary behaviour is an interesting observation. We attribute this behaviour to the precise feature extraction of the SSL-based frontend, which can capture subtle noise patterns, spectral smoothing, temporal artefacts, and phase distortions that are more challenging for the simpler, convolution-based networks to detect. 

\begin{figure*}
    \centering
    \begin{subfigure}{\textwidth}
        \centering
        \includegraphics[width=0.8\linewidth]{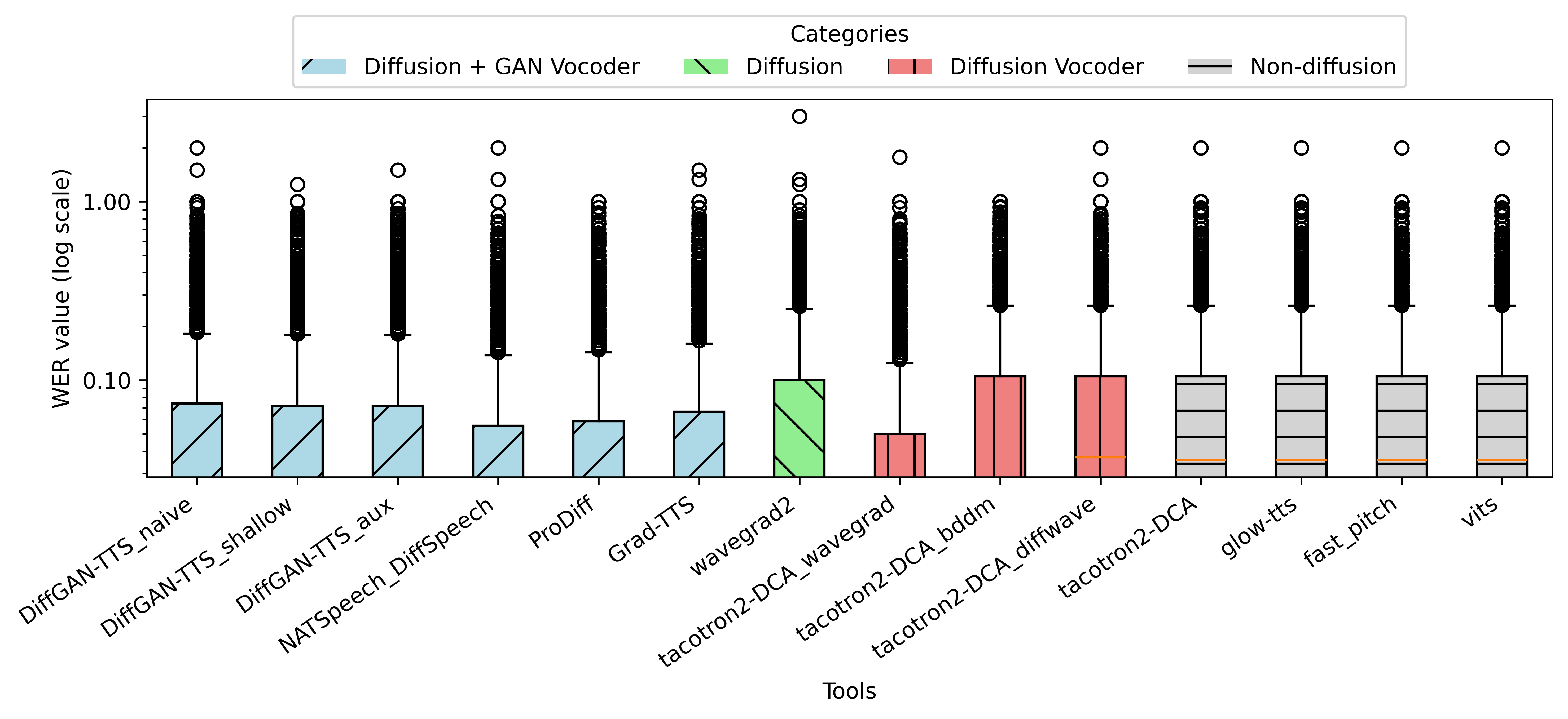}
        \caption{WER. \textit{Y-axis uses log scale for better readability.}}
        \label{fig:quality_plots_wer}
    \end{subfigure}
    \vfill
    \begin{subfigure}{\textwidth}
        \centering
        \includegraphics[width=0.8\linewidth]{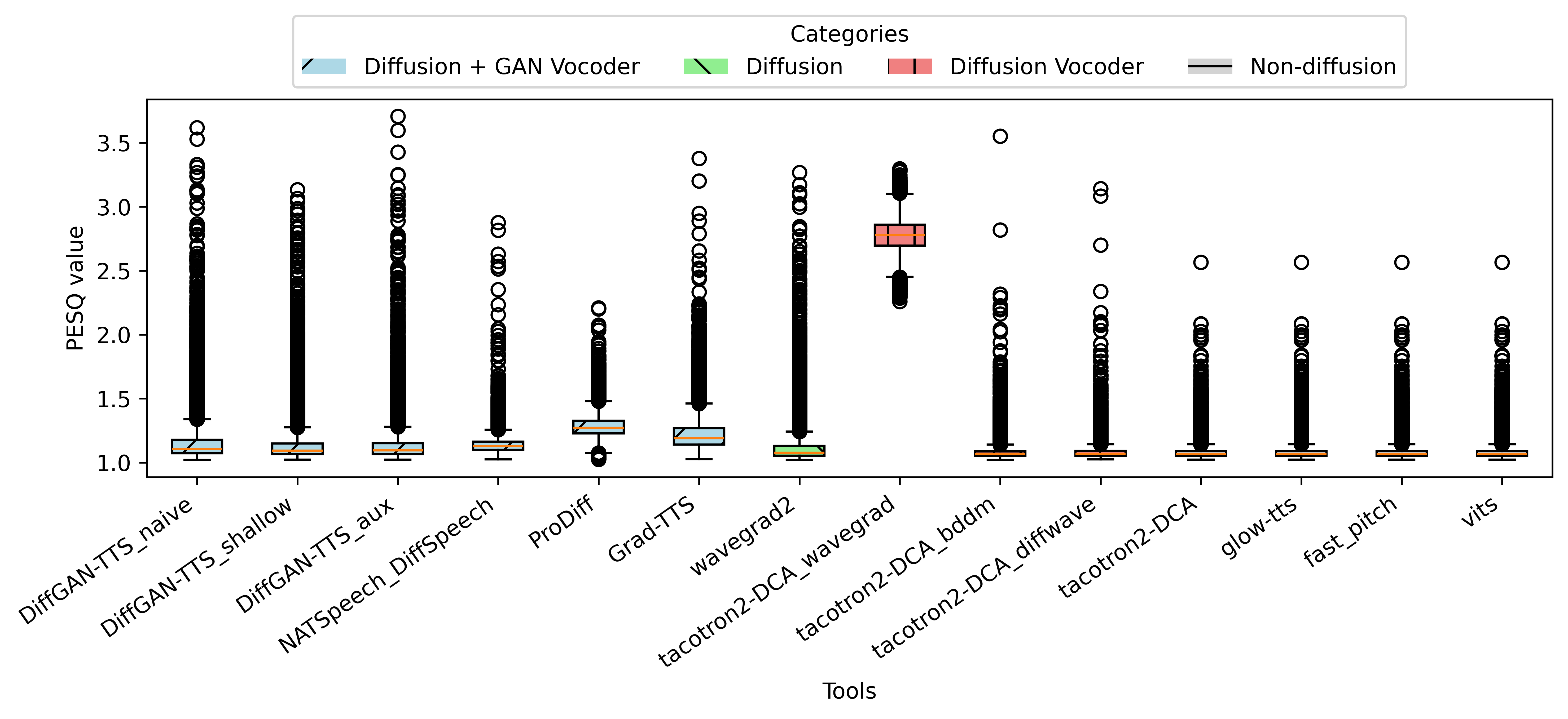}
        \caption{PESQ}
        \label{fig:quality_plots_pesq}
    \end{subfigure}
    \vfill
    \begin{subfigure}{\textwidth}
        \centering
        \includegraphics[width=0.8\linewidth]{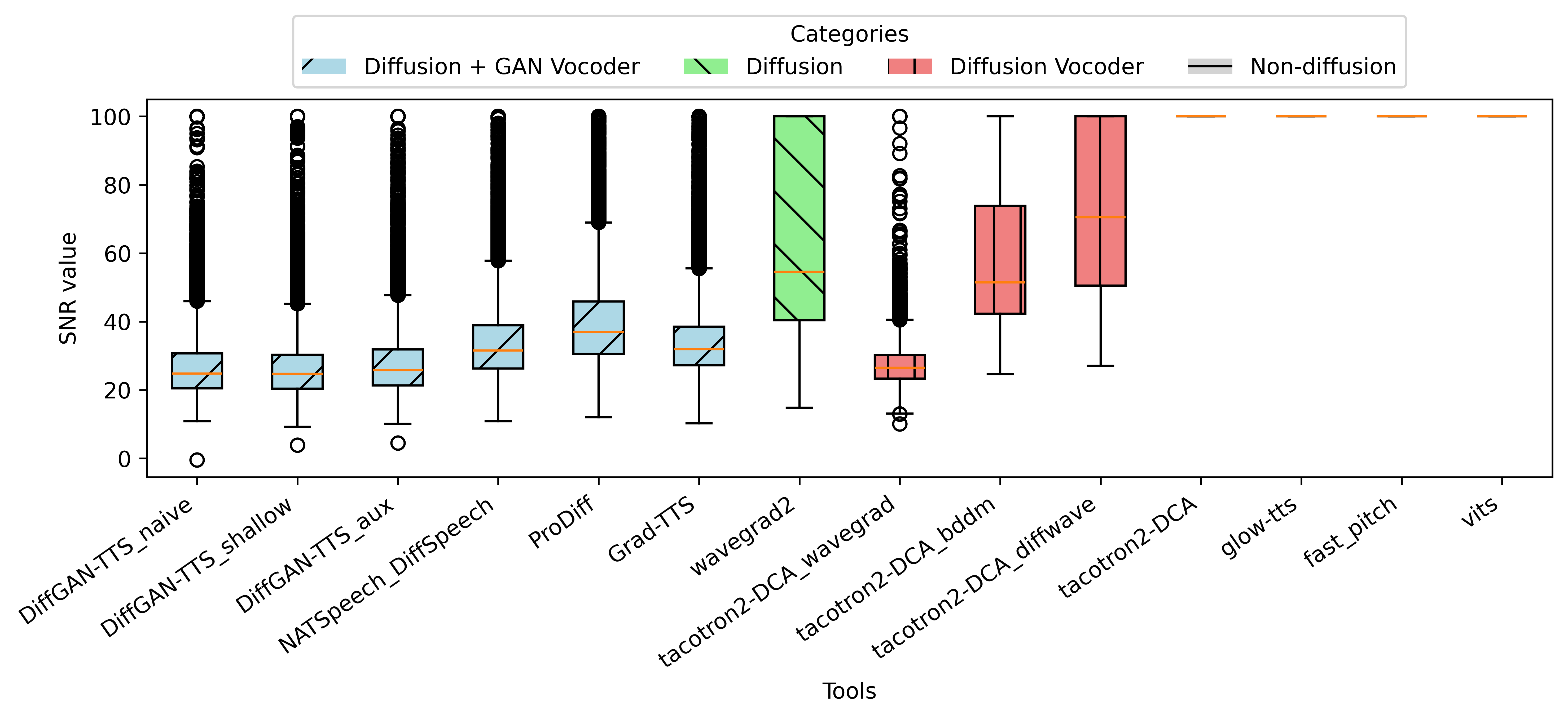}
        \caption{SNR}
        \label{fig:quality_plots_snr}
    \end{subfigure}
    \caption{Visualization of quality metrics for used tools.}
    \label{fig:quality_plots}
\end{figure*}

\subsection{Speech quality}

The quality assessment measured Speaker similarity, WER, PESQ and SNR metrics. Firstly, the speaker similarity was consistent across all tools and close to 1.0. All of the tools thus reproduced the original speaker convincingly. The remaining metrics are visualized in Fig.~\ref{fig:quality_plots}. All the tools convert the text to spoken content with no problems, with low WER rates (Fig.~\ref{fig:quality_plots_wer}). Similarly, PESQ (Fig.~\ref{fig:quality_plots_pesq}) remains quite similar for all tools and indicates good quality of synthesized speech; the only exception is diffusion-vocoder wavegrad which shows higher variability of observed PESQ scores. Finally, the greatest differences are seen regarding SNR (Fig.~\ref{fig:quality_plots_snr}), where the non-diffusion vocoders reach high values, indicating good quality, and all other synthesizers are spread across the whole range of scores. The diffusion-synthesized samples with the GAN vocoder show the highest amount of noise. 

Finally, there is no significant difference in the synthesis speed based on the average observed RTF value for non-diffusion (0.073) and diffusion synthesizers (0.064). The only exception is WaveGrad2, with an RTF of almost 17.50, which means that the synthesis of one minute of speech takes almost 18 minutes.

\section{Discussion}

The landscape of diffusion synthesizers is diverse, but their usability is still limited due to the reliance on pretrained models, mostly available for LJspeech. To improve their applicability, developing and training these models on a broader range of datasets is essential. Expanding the availability of pretrained models would make diffusion synthesizers more versatile and facilitate more comprehensive security assessments of these systems.

Contrary to our initial concerns, the impact of diffusion models on detection methods is less significant than anticipated, which is a positive outcome. This finding alleviates worries about these models posing a serious threat to deepfake detection accuracy. While some synthesizer-detector pairs may experience a slight decline in accuracy, this decrease is not substantial enough to raise major concerns. Furthermore, combining multiple detectors will likely provide robust detection capabilities for identifying these samples.

An interesting observation is the prevalent use of diffusion synthesizers with GAN vocoders. Moreover, the diffusion process introduces additional noise into the synthesized speech, as lower SNR values indicate. This phenomenon could be a promising avenue for future research to identify diffusion-synthesized speech more effectively.

The primary contribution of this paper is to deliver a novel dataset. The supplementary analyses offer context and insight into the dataset's characteristics and the rationale behind its creation. While these analyses may be limited in scope, this does not detract from their value, as the primary focus remains publishing the data.

\subsection{Limitations}

The primary limitation of this work is the use of only the LJspeech dataset, which lacks diversity in speakers. Consequently, the results observed in deepfake speech detection may be biased due to this limited variability. 

Additionally, the results presented in this study are based on proof-of-concept (PoC) experiments utilizing a limited number of detectors. As a result, these initial observations necessitate further validation through extensive testing with a broader array of detectors to ensure robustness and accuracy. Despite the preliminary nature of these findings, the primary contribution of this research is the development and provision of a comprehensive dataset. Therefore, the limited scope of the current results does not undermine the overall significance and utility of the study.

\section{Conslusions}

In conclusion, the detection of diffusion-based deepfakes demonstrates a level of similarity to that of non-diffusion deepfakes. While there are nuanced differences, the general efficacy of detection remains consistent across both types of tools, exhibiting strengths and weaknesses.

Our findings indicate that re-vocoding samples using diffusion vocoders does not significantly impact detection outcomes. The detection effectiveness largely depends on the specific architecture and settings of the detection model.

Furthermore, the audio quality generated by diffusion-based tools is comparable to that of non-diffusion tools. However, diffusion-based methods introduce noise into the final recordings, as evidenced by lower Signal-to-Noise Ratios (SNR).

\section*{Acknowledgements}

This work was supported by the national project NABOSO: Tools To Combat Voice DeepFakes (with code VB02000060), funded by the Ministry of the Interior of the Czech Republic and the Brno University of Technology internal project FIT-S-23-8151.

\bibliographystyle{IEEEtran}
\bibliography{diffusion_dataset_007}

\end{document}